\documentstyle[12pt, a4]{article}
\begin{document}
\author{B. G\"{o}n\"{u}l and M. Ko\c{c}ak
\and Department of Engineering Physics, Faculty of Engineering,
\and University of Gaziantep, 27310 Gaziantep -T\"{u}rkiye}
\title{An Approach To Potential Scattering}
\date{}
\maketitle
\begin{abstract}

Recently developed time-independent bound-state perturbation
theory is extended to treat the scattering domain.The changes in
the partial wave phase shifts are derived explicitly and the
results are compared with those of other methods.

\end{abstract}



\section{Introduction}

In our previous papers [1-6] published recently, a
time-independent novel perturbation theory has been developed in
the bound state domain, which is non-perturbative, self-consistent
and systematically improvable,  and used to treat successfully
significant problems in different fields of physics. Gaining
confidence from these applications, we aim through the present
work to show that similar techniques can also be used in the
continuum.

In the next section we summarize the main ideas of our approach.
The extension of the model for scattering states and the
relationship to some other perturbation approaches are discussed
in Section 3. The paper ends with a brief summary and concluding
remarks.

\section{The Model}
Let us start with a brief introduction of the formalism to remind
the compact form of the method, which would provide an easy access
of the scheme to understanding of the treatments in the continuum.
For the consideration of spherically symmetric potentials, the
corresponding Schr\"{o}dinger equation in the bound state domain
for the radial wave function has the form $(\hbar=2m=1)$

\begin{equation}
\frac{\Psi''_{n}(r)}{\Psi_{n}(r)}=[V(r)-E_{n}],
~~~V(r)=\left[V_{0}(r)+\frac{\ell(\ell+1)}{r^{2}}\right]+\Delta{V(r)},~~~~n=0,1,2,...,
\end{equation}
where $V_{0}$  is an exactly solvable unperturbed potential
together with the angular momentum barrier while $\Delta V$ is a
perturbing potential. Expressing the wave function $\Psi_{n}$ as a
product
\begin{equation}
\Psi_{n}(r)=\chi_{n}(r)\phi_{n}(r),
\end{equation}
in which $\chi_{n}$ is the known normalized eigenfunction of the
unperturbed Schr\"{o}dinger equation whereas $\phi_{n}(r)$ is a
moderating function corresponding to the perturbing potential.
Substituting (2) into (1) yields
\begin{equation}
\left(\frac{\chi''_{n}}{\chi_{n}}+\frac{\phi''_{n}}{\phi_{n}}+2\frac{\chi'_{n}}{\chi_{n}}\frac{\phi'_{n}}{\phi_{n}}\right)=V-E_{n}.
\end{equation}
Instead of setting the functions $\chi_{n}$  and $\phi_{n}(r)$, we
will set their logarithmic derivatives
\begin{equation}
W_{n}=-\frac{\chi'_{n}}{\chi_{n}},
~~~\Delta{W_{n}}=-\frac{\phi'_{n}}{\phi_{n}}
\end{equation}
which leads to
\begin{equation}
\frac{\chi''_{n}}{\chi_{n}}=W_{n}^{2}-W'_{n}=\left[V_{0}(r)+\frac{\ell(\ell+1)}{r^{2}}\right]-\varepsilon_{n},
\end{equation}
where $\varepsilon_{n}$  is the eigenvalue of the exactly solvable
unperturbed potential, and
\begin{equation}
\left(\frac{\phi''_{n}}{\phi_{n}}+2\frac{\chi'_{n}}{\chi_{n}}\frac{\phi'_{n}}{\phi_{n}}\right)=
\Delta {W^{2}_{n}}-\Delta {W'_{n}}+2W_{n}\Delta {W_{n}}=\Delta
{V(r)}-\Delta\varepsilon_{n}
\end{equation}
in which $\Delta\varepsilon_{n}$ is the energy value for the
perturbed potential leading to
$E_{n}=\varepsilon_{n}+\Delta\varepsilon_{n}$. If the whole
potential, involving the perturbing piece $\Delta{V}$, can be
analytically solvable, then Eq.(1) through (5) and (6) reduces to
\begin{equation}
(W_{n}+\Delta {W_{n}})^{2}-(W_{n}+\Delta {W_{n}})'=V-E_{n},
\end{equation}
which is known as the usual supersymmetric quantum mechanical
treatment \cite{cooper} in the literature.

However, if the whole potential has no analytical solution as the
case considered in this Letter, which means Eq.(6) cannot be
exactly solvable for $\Delta{W}$, then one can expand the
functions in terms of the perturbation parameter $\lambda$,
\begin{equation}
\Delta{V(r;\lambda)}=\sum^{\infty}_{N=1}\lambda^{N}\Delta{V_{N}(r)},
~~\Delta{W_{n}(r;\lambda)}=\sum^{\infty}_{N=1}\lambda^{N}\Delta{W_{nN}(r)},\nonumber
~~\Delta\varepsilon_{n}(\lambda)=\sum^{\infty}_{N=1}\lambda^{N}\Delta\varepsilon_{nN}
\end{equation}
where $N$ denotes the perturbation order. Substitution of the
above expansion into Eq.(6) and equating terms with the same power
of $\lambda$ on both sides yields up to for instance
$O(\lambda^{3})$
\begin{equation}
2W_{n}\Delta {W_{n1}}-\Delta {W'_{n1}}=\Delta
{V_{1}}-\Delta\varepsilon_{n1},
\end{equation}
\begin{equation}
\Delta{W^{2}_{n1}}+2W_{n}\Delta{W_{n2}}-\Delta{W'_{n2}}=\Delta{V_{2}}-\Delta\varepsilon_{n2}
\end{equation}
\begin{equation}
2(W_{n}\Delta{W_{n3}}+\Delta{W_{n1}}\Delta{W_{n2}})-\Delta{W'_{n3}}=\Delta{V_{3}}-\Delta{\varepsilon_{n3}}
\end{equation}
Eq.(6)and its expansion through Eqs.(9-11) give a flexibility for
the easy calculations of the perturbative corrections to energy
and wave functions for the $\textit{nth}$ state of interest
through an appropriately chosen perturbed superpotential. It has
been shown [1-6] that this feature of the present model leads to a
simple framework in obtaining the corrections to all states
without using complicated mathematical procedures.

\section{Application to the scattering domain}
It is well known that there are many scattering problems in which
the interaction between the projectile and the target decomposes
naturally into two parts $(V=V_{0}+\Delta{V})$. This division is
especially useful if the scattering wave function under the action
one part can be obtained exactly $(V_{0})$, while the effect of
the other $(\Delta{V})$ can be treated in some approximation as in
the present formalism.

For simplicity, we here confine ourselves to $s-$wave scattering
from a potential which is assumed that vanishes beyond a finite
radius $R$. The associated total wavefunction behaves at large
distances
\begin{equation}
\Psi(r)=\frac{1}{k}\sin(kr+\delta) ,~~~r \geq R ,
\end{equation}
where $\delta$ is the $s-$wave phase shift.

Our present treatment of scattering has concerned itself primarily
with determining how the solutions of the free Schr\"{o}dinger
equation are affected by the presence of the interaction. Within
the framework of the present formalism we suppose that the
solutions of Eq.(5) are known, or are easily found, to give the
corresponding phase shift $\delta_{0}$. Considering the expansion
$\delta=\delta_{0}+\lambda\delta_{1}+\lambda^{2}\delta_{2}+...$,
as in Eq.(8), we aim here to derive explicitly solvable and easily
accessible expressions for the phase shift contributions at
successive perturbation orders.

\subsection{First-order phase shift correction}
Keeping in mind Eq.(12) and considering the discussion in Section
2, at the first perturbation order one has
\begin{equation}
(W+\lambda\Delta{W_{1}})=-k\cot(kr+\delta_{0}+\lambda\delta_{1}),~~~W_{n}=-\frac{\chi'}{\chi}=-k\cot(kr+\delta_{0}),
\end{equation}
from where the superpotential relating to the perturbing
interaction
\begin{equation}
\Delta{W_{1}(r)}=\frac{k\delta_{1}}{\sin^{2}(kr+\delta_{0})}~,
\end{equation}
is obtained assuming that
$\sin\lambda\delta_{1}\cong\lambda\delta_{1}$ and
$\cos\lambda\delta_{1}\cong1 $.

In the second step, one needs to employ Eq. (9) to arrive at
another expression for $\Delta{W_{1}}$. Rearranging the terms,
$\Delta{W'_{1}}-2W\Delta{W_{1}}=(\Delta\varepsilon_{1}-\Delta{V_{1})}$
and multiply both sides by the integrating factor
$\exp(-2\int^{r}_{0}W(z)dz)$, which is the square of the
unperturbed wave function $\chi^{2}(r)$ through Eq.(4), one obtain
\begin{equation}
\frac{d}{dr}\left[\chi^{2}(r)\Delta{W_{1}(r)}\right]=\chi^{2}(r)(\Delta\varepsilon_{1}-\Delta{V_{1}}).
\end{equation}
The integration, and the remove of $\Delta{\varepsilon_{1}}$  term
due to the consideration of elastic scattering process here,
yields
\begin{equation}
\Delta{W_{1}(r)}=-\frac{1}{\chi^{2}(r)}\int^{r}_{0}\chi^{2}(z)
\Delta{V_{1}(z)}dz.
\end{equation}
As $\chi=\frac{1}{k}\sin(kr+\delta_{0})$ in the asymptotic region,
comparison of Eqs.(14) and (16) reproduces the first-order change
in the phase shift
\begin{equation}
\delta_{1}=-k\int^{\infty}_{0}\chi^{2}(r) \Delta{V_{1}(r)}dr.
\end{equation}
If necessary, the corresponding change in the wavefunction can
easily be obtained by the substitution of Eq.(16) into (4),
$\phi_{1}=\exp(-\int\Delta{W_{1}})$. For the reliability of the
present expression obtained, Eq (17), one may compare it with that
reproduced by other methods. For example, in the limiting case
where the unperturbed potential vanishes, the unperturbed $s-$wave
function is reduced to a plane wave $\chi(r)=\sin(kr)/k$, and the
first-order change in the phase shift becomes
\begin{equation}
\delta_{1}=-\frac{1}{k}\int^{\infty}_{0}\sin^{2}(kr)
\Delta{V_{1}(r)}dr
\end{equation}
which is just the first Born approximation for the phase shift
\cite{thaler}. In addition, the well known expression for $s-$wave
scattering amplitude by the two-potential formula in scattering
theory \cite{thaler},
\begin{equation}
f_{1}=-e^{2i\delta_{0}}\int^{\infty}_{0}\chi^{2}(r)
\Delta{V_{1}(r)}dr
\end{equation}
where the phase factor in front of the integration arises because
of the standing wave boundary conditions, justifies once more our
result since $f_{1}=-e^{2i\delta_{0}}\delta_{1}/k$ and, equating
this to the above equation leads immediately to Eq.(17).

The present result has a widespread applicability, which may also
be used in the treatment of scattering length problems. At
low-energy limit, the phase shift is related to the scattering
length $\delta_{k\rightarrow{0}}\rightarrow{-ka}$ where
${a}={a_{0}}+\lambda{a_{1}}+\lambda^{2}{a_{2}}+...$ may be
expanded in a perturbation series similar to the phase shift.
Outside the range of the potential, the unperturbed wave function
behaves as $\chi\rightarrow(r-a_{0})$. Thus, the first correction
to the scattering length is
\begin{equation}
a_{1}=\lim_{r\rightarrow\infty}\left[\int^{r}_{0}(z-a_{0})^{2}\Delta{V_{1}(z)}dz\right]
\end{equation}
which can be calculated for a given $\Delta{V_{1}}$. The
scattering length has an important physical significance. In the
low-energy limit only the $s-$wave makes a nonzero contribution to
the cross section, so that the angular distribution of the
scattering is spherically symmetric and the total cross section is
$4\pi(a_{0}+\lambda{a_{1}}+...)^{2}$. This is also exactly the
result obtained in most textbooks for the low-energy scattering of
a hard sphere of radius Thus the scattering length is the
effective radius of the target at zero energy.

As a last example, consider the case of the angular momentum
barrier as the unperturbed potential $V_{0}=\ell(\ell+1)/r^{2}$
that produces $\left[rj_{\ell}(kr)\right]$ with a phase shift
$\delta_{0}=-\ell\pi/2$. For a trivial perturbation let us choose
$\Delta{V_{1}}=\lambda/r^{2}$, due to which the angular momentum
is slightly perturbed
$\overline{\ell}\approx\ell+\lambda/(2\ell+1)+O(\lambda^{2})$.
Therefore the phase shift correction at first-order is
$\delta_{1}=-\pi/2(2\ell+1)$. Again, this exact result confirms
the reliability of Eq.(17).

\subsection{Second-order phase shift correction}
To solve Eq.(10) for $\Delta{W_{2}}$ we mimic the preceding
calculation. The integration factor is the same. In fact,
examining Eqs.(9) and (10), the only difference is that  the
quantity $\Delta{V_{1}}-\Delta\varepsilon_{1}$ is replaced by
$\Delta{V_{2}}-\Delta{W^{2}_{1}}-\Delta\varepsilon_{2}$. As
$\Delta\varepsilon_{2}$ term is zero due to the process of
interest, $\Delta{W_{2}}$ is thus
\begin{equation}
\Delta{W_{2}(r)}=-\frac{1}{\chi^{2}(r)}\int^{r}_{0}\chi^{2}(z)
\left[\Delta{W^{2}_{1}(z)}-\Delta{V_{2}(z)}\right]dz.
\end{equation}
Bearing in mind that $\chi=\frac{1}{k}\sin(kr+\delta_{0})$ for the
region $r\geq{R}$, the second-order expansion in the
superpotential similar to Eq.(13) provides another expression for
$\Delta{W_{2}}$ which is
\begin{equation}
\Delta{W_{2}(r)}=k\delta_{1}^{2}\frac{\cot(kr+\delta_{0})}{\sin^{2}(kr+\delta_{0})}+\frac{k\delta_{2}}{\sin^{2}(kr+\delta_{0})}
\end{equation}
Comparison of Eqs.(21) and (22), together with  the substitution
of (14) in (21), leads to an auxiliary function for the second
order phase shift correction,
\begin{equation}
\delta_{2}(r)=-\frac{1}{k}\int^{r}_{0}\Delta{V_{2}(z)}\sin^{2}(kz+\delta_{0})
dz
+k\delta_{1}^{2}\int^{r}_{0}\frac{dz}{\sin^{2}(kz+\delta_{0})}-\delta_{1}^{2}\cot(kr+\delta_{0}),
\end{equation}
where a singularity appears in the second integral at $z=0$. This
problem can be circumvented by replacing the lower limit of the
integral with $R$. Assuming $\Delta{V}=\Delta{V_{1}}$ as in
realistic problems of nuclear physics, which means that
$\Delta{V_{2}}=0$, the $r-$dependent phase shift correction in the
second-order is given in the form of
\begin{equation}
\delta_{2}(r)=\delta_{1}^{2}\cot(kR+\delta_{0})-2\delta_{1}^{2}\cot(kr+\delta_{0}).
\end{equation}
As an alternative treatment, which leads to a concrete comparison,
one can go back to Eq.(21) and split $\chi^{2}\Delta{W_{1}^{2}}$
term in two parts as $(\chi^{2}\Delta{W_{1}})(\Delta{W_{1}})$
allowing to invoke Eq.(16). In this case the comparison of the
result with the expansion in (22) gives
\begin{equation}
\delta_{2}=-k\int^{\infty}_{0}\chi^{2}(r)\Delta{V_{1}(r)}dr
\int^{r}_{R}\frac{dz}{\chi^{2}(z)}
\left[\int^{R}_{z}\chi^{2}(y)\Delta{V_{1}(y)}dy-\frac{\delta_{1}}{k}\right]+\delta_{1}^{2}\cot(kR+\delta_{0})
\end{equation}
which is in agreement with the work in \cite{milward}. In
addition, the use of (17) in (24) transforms it into Eq. (25).
Furthermore, the reader is reminded that the second Born
approximation for the phase shift can be most easily derived using
the variable phase equation approach \cite{calegero},
\begin{equation}
\delta_{2}=2k^{2}\int^{\infty}_{0}\chi^{2}(r)\Delta{V_{1}(r)}\cot(kr)dr\int^{r}_{0}\chi^{2}(y)\Delta{V_{1}(y)}dy
\end{equation}
which, in the light of Eq. (15), is the same result as we find
from Eq (25), by putting $\delta_{0}=0$ . Higher order terms can
also be evaluated in the same manner.

\section{Concluding Remarks}
The recently introduced time-independent perturbation theory has
been successfully extended from the bound state region to the
scattering domain. For the clarification, the work has been
carried out with the consideration of $s-$wave scattering only.
However, generalization of the formalism to higher partial waves
in the scattering domain does not cause any problem. The inclusion
of the centrifugal barrier contribution in the effective potential
for instance leads to the replacement of the $s-$wave phase shift
with $\delta_{\ell}-\ell\pi/2$ due to the related wave function
$\chi(r)=\sin(kr+\delta_{\ell}-\ell\pi/2)/k$ in the asymptotic
region, supposing both the unperturbed and perturbed potentials
vanish at a large $r>R_{1}$ which means that in the region
$R_{1}<r\leq{R}$ there is then only the centrifugal barrier
contribution. This inclusion requires simply to repeat the present
calculations for the replacement in the phase shift.

It should be stress that, anything that can be achieved from the
present formalism must also be obtainable from the works [9,10] in
the literature. For instance, considering the bound  state region,
Bender's formalism \cite{bender} can be simplified by introducing
the auxiliary function $F_{N}(r)$  such that the whole wave
function $\Psi_{N}(r)=\chi(r)F_{N}(r)$ where denotes the
perturbation order. The first-order correction can then be written
as
$\frac{d}{dr}\left[\chi^{2}\frac{dF}{dr}\right]=(\Delta{V_{1}}-\Delta\varepsilon_{1})\chi^{2}$
which corresponds exactly to the present treatment by Eq. (15)
when we identify $\Delta{W_{1}}=dF/dr$. The higher order
calculations can be linked to ours in the similar manner. Whereas,
the works of Milward and Wilkin \cite{milward} may be related to
the present formalism in both domain, the bound and scattering
region by making a relation between their probability density
distributions/derivatives and our $\Delta{W}$ functions, such as
$\Delta{W_{0}}=-P_{0}'/2P_{0}$ at the zeroth order,
$\Delta{W_{1}}=(-P_{1}/2P_{0})'$ at the first order and
$\Delta{W_{2}}=(-P_{2}/2P_{0})'$  at the second order etc.
Nevertheless, the present technique provides a clean and explicit
route for the calculations without tedious and cumbersome
integrals.

The energy variation of the scattering wave function and phase
shift can also be studied by perturbing in the energy. We wish to
stress that all these effects depend purely upon the perturbation
and the unperturbed wave function; explicit knowledge of the
unperturbed potential is not necessary. This exposition will be
deferred to a later publication.

\end{document}